\documentclass[final,1p,times]{elsarticle} 

\usepackage{graphicx}
\usepackage{amssymb} 
\usepackage{amsthm} 

\newcommand{\pt}{$p_{{\mathrm T}}$~}

\newcommand{\jp}{J$/\psi$~}
\newcommand{\raa}{$R_{\mathrm{AA}}$~}
\newcommand{\sq}{$\sqrt{s_{\mathrm{NN}}}$}
\newcommand{\ncoll}{$N_{\mathrm{coll}}$}
\newcommand{\ee}{e$^+$+e$^-$~}
\newcommand{\cc}{${\rm c}\bar{{\rm c}}$~}

\journal{Nuclear Physics A} 

\begin{document}

\begin{frontmatter} 

\title{\jp nuclear modification factor at mid-rapidity in Pb-Pb collisions at \sq=2.76~TeV}

\author{Ionut-Cristian Arsene for ALICE Collaboration$^1$}
\fntext[col1] {A list of members of the ALICE Collaboration and acknowledgements can be found at the end of this issue.}
\address{{\it Research Division and ExtreMe Matter Institute EMMI, GSI Helmholtzzentrum f\"{u}r 
         Schwerionenforschung, Darmstadt, Germany}}

\begin{abstract} 
We report on the \jp nuclear modification factor \raa at mid-rapidity ($|y|<0.9$) in Pb-Pb collisions at \sq=2.76~TeV
measured by ALICE. \jp candidates are reconstructed using their \ee decay channel. The kinematical coverage
extends to zero transverse momentum allowing the measurement of integrated cross sections.
We show the centrality dependence of the \jp \raa at mid-rapidity compared to the results from PHENIX at
mid-rapidity and ALICE results at forward-rapidity. We also discuss comparisons to calculations
from theoretical models.
\end{abstract} 

\end{frontmatter} 


Based on considerations in Quantum Chromo Dynamics (QCD) it is commonly believed
that during the lifetime of the fireball of a heavy-ion collision at the modern high energy accelerators,
a new phase of hadronic matter may be formed. This state, called quark-gluon plasma (QGP), consists
of deconfined quarks and gluons at very high temperature or energy density. 
Charmonia are strongly bound meson states of \cc pairs. Due to their large mass, the charm quark pairs can be 
created only in the first instants of the collision, during the partonic stage, which enables them
to probe the entire evolution of the nuclear fireball.
The suppression of charmonium states via the color screening effect was one of the first proposed signals
to test the QGP \cite{matsui86}. 
Experimental results from heavy-ion collisions at SPS and RHIC exhibit very similar and small values of the nuclear modification
factor \raa \cite{na50,phenix2007}.
Although the observation of such a \jp suppression
at two different energies is striking, the results need to be interpreted very carefully since there 
are additional energy dependent effects involved in 
charmonium production \cite{brambilla2011}, {\it e.g.} cold nuclear matter effects. Another important 
effect is the feed-down from higher mass charmonium states
and from beauty hadron decays which are often not taken into account due to experimental limitations. 
Furthermore, when the charm-quark density is high enough, charmonium states may be (re)created
at the QGP state breakup \cite{pbm00, andronic07} or during its evolution \cite{zhao2011,liu2009}. 
This effect is expected to give a significant contribution
to the \jp yields at the LHC energies. In the following we will give a brief report on the latest ALICE
results on \jp production at mid-rapidity in Pb-Pb collisions at \sq=2.76~TeV.

A detailed description of the ALICE experimental setup is provided in \cite{aliceJINST}.
The present analysis is based on 30 million Pb-Pb collisions at \sq=2.76~TeV which, compared to the
previous results shown in \cite{jensHP2012} includes data from the 2011 LHC Pb-Pb run in addition.
At mid-rapidity, the \jp particles are measured using their di-electron decay channel. The electrons
are reconstructed using a system of tracking and particle identification detectors which cover the pseudo-rapidity range
$|\eta|<0.9$. Our \jp coverage at mid-rapidity ($|y|<0.9$) extends to zero transverse momentum (\pt).
Our present measurement does not allow to separate the feed-down contributions, so in the following
we always refer to inclusive \jp measurements. ALICE also measures \jp at forward rapidity. For a detailed
account on the forward rapidity measurements refer to \cite{robertaQM2012,hongyanQM2012}.

\begin{figure}[!tp]
\begin{center}
\begin{tabular}[htb]{cc}
\includegraphics[width=0.35\textwidth]{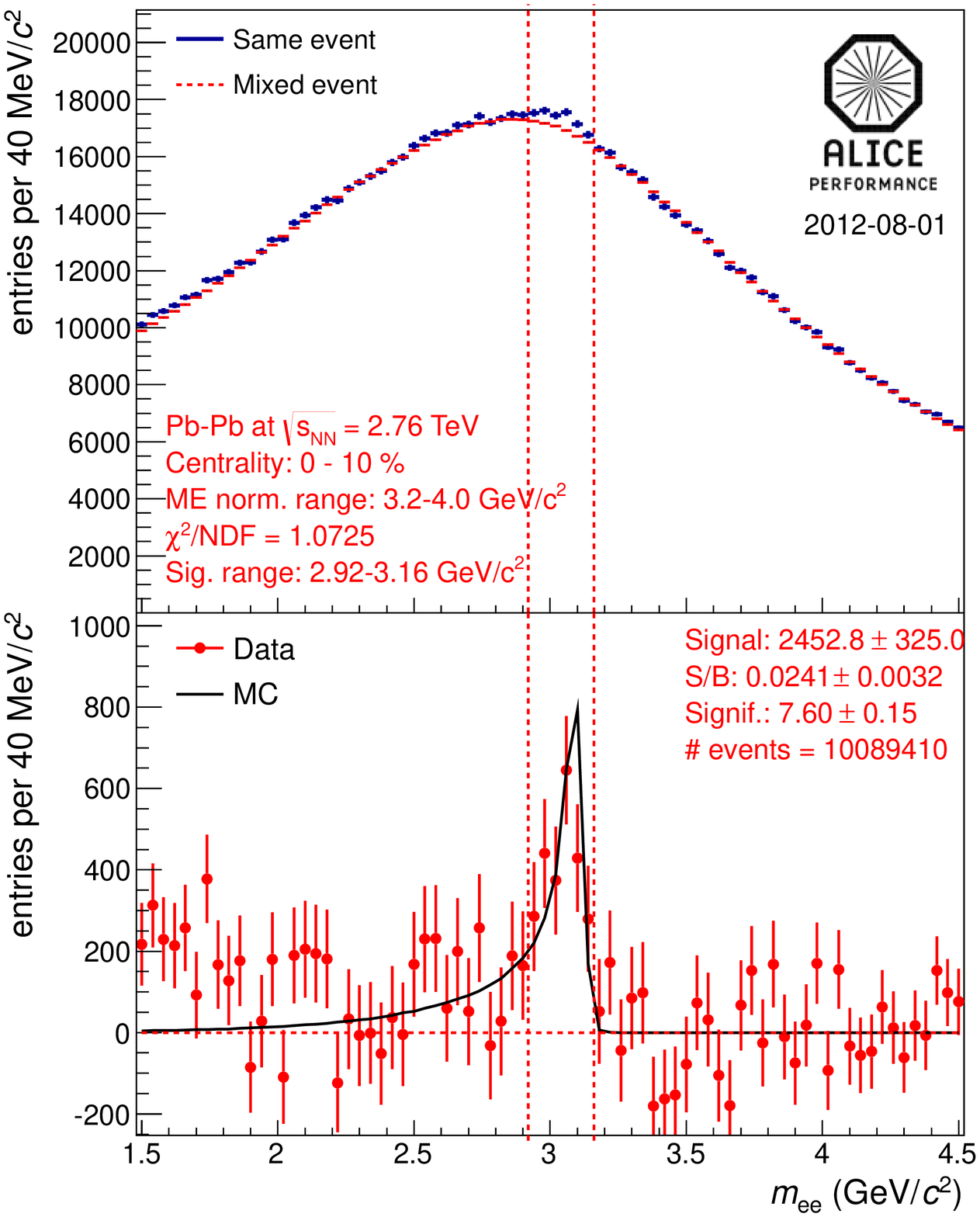}
&
\includegraphics[width=0.35\textwidth]{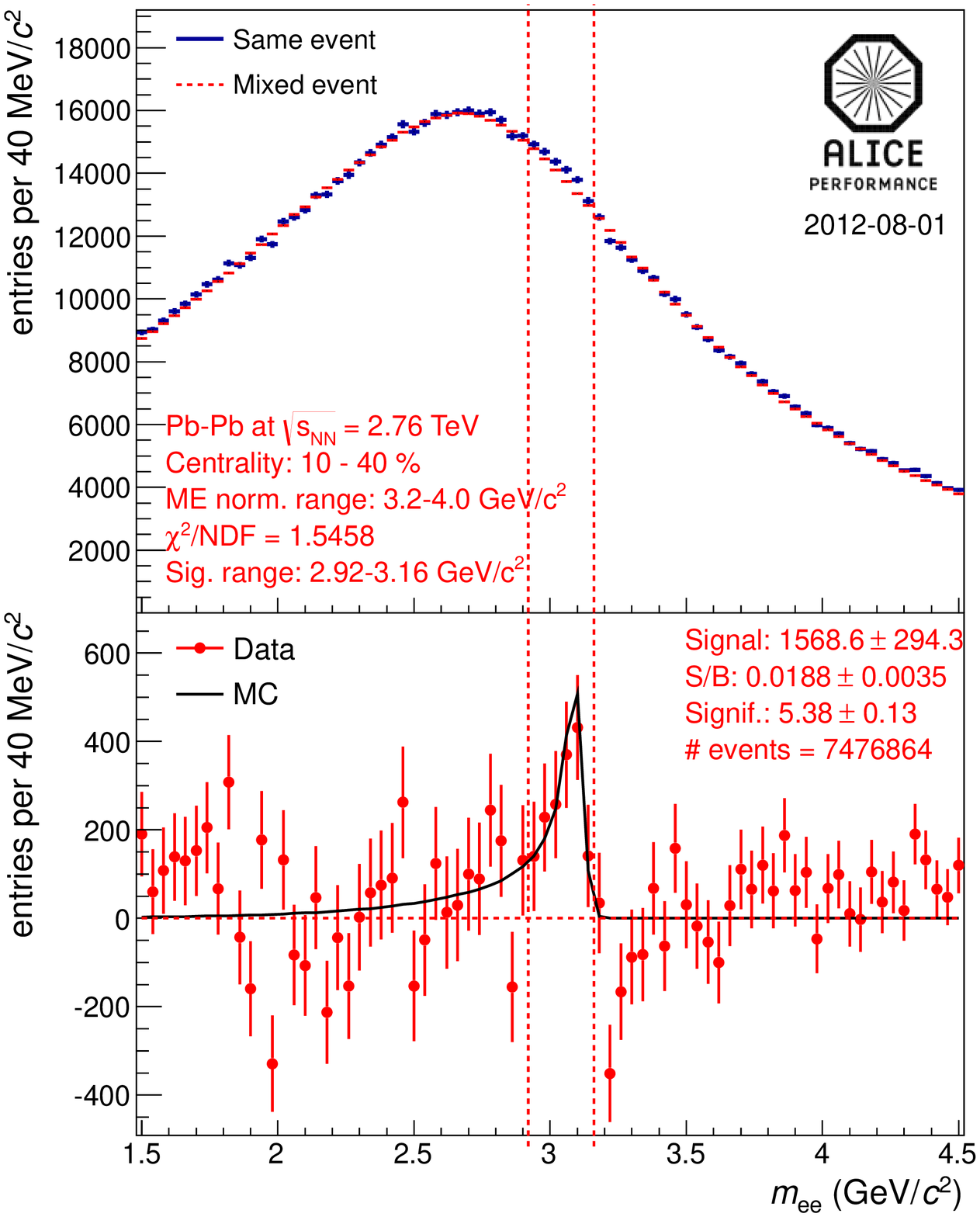} \\
\end{tabular}
\end{center}
\caption{Invariant mass distribution in 0-10\% (left) and 10-40\% (right) Pb-Pb collisions.}
\label{fig:invmass}
\end{figure}

The \jp signal is extracted using the invariant mass distribution constructed from opposite-sign (OS) pairs
of electron candidates (see Fig.\ref{fig:invmass}). The dominant contribution in the OS invariant mass distribution
is coming from uncorrelated pairs which are subtracted using the event mixing technique. The mixed event background
distribution is normalized to match the same-event OS distribution in the mass range $3.2<m<4.0$~GeV/$c^2$, where the
\jp signal is expected to vanish.
A good matching between the same event and mixed event distributions is observed over a broad mass range 
meaning that the contribution to the \jp
signal from the \ee continuum ({\it e.g.} semi-leptonic charm and beauty decays, Drell-Yan, etc.) 
is small and can be included in the signal extraction systematic uncertainty.
The bottom panels of Fig.\ref{fig:invmass} show the background-subtracted invariant mass distributions in comparison
to the \jp signal shape as obtained from our Monte-Carlo (MC) simulations of the ALICE detector setup. The long tail toward
lower masses is due to the electron energy loss in the detector material via bremsstrahlung and to the radiative decay channel, 
\jp$\rightarrow$\ee$+\gamma$, which generates a soft photon that is not reconstructed.  
The raw \jp signal is extracted using bin counting in the mass range $2.92-3.16$~GeV/$c^{2}$. The extrapolation to the full
mass range is included in the efficiency correction.

The raw \jp signal was corrected for acceptance and efficiency using simulated Pb-Pb collisions from HIJING 
enriched with primary \jp.
All particles from these collisions were transported through a GEANT simulation of the ALICE setup. 
The input rapidity and \pt distributions of the embedded \jp was obtained
from an interpolation of RHIC, Fermilab and LHC data \cite{bossu11}. The uncertainty on the integrated yields due to unknown
\pt and rapidity distributions was found to be $\approx$~2\%. The efficiency factors obtained after applying the same
cuts as in the data analysis amount to $\approx$~3.2\%, 7.4\% and 8.0\% in the 0-10\%, 10-40\% and 40-80\% centrality classes, 
respectively. The large difference in efficiency between central and semi-central collisions is due to the different cut 
strategies chosen to improve the significance of the signal.

\begin{figure}[!tp]
\begin{center}
\begin{tabular}[htb]{cc}
\includegraphics[width=0.50\textwidth]{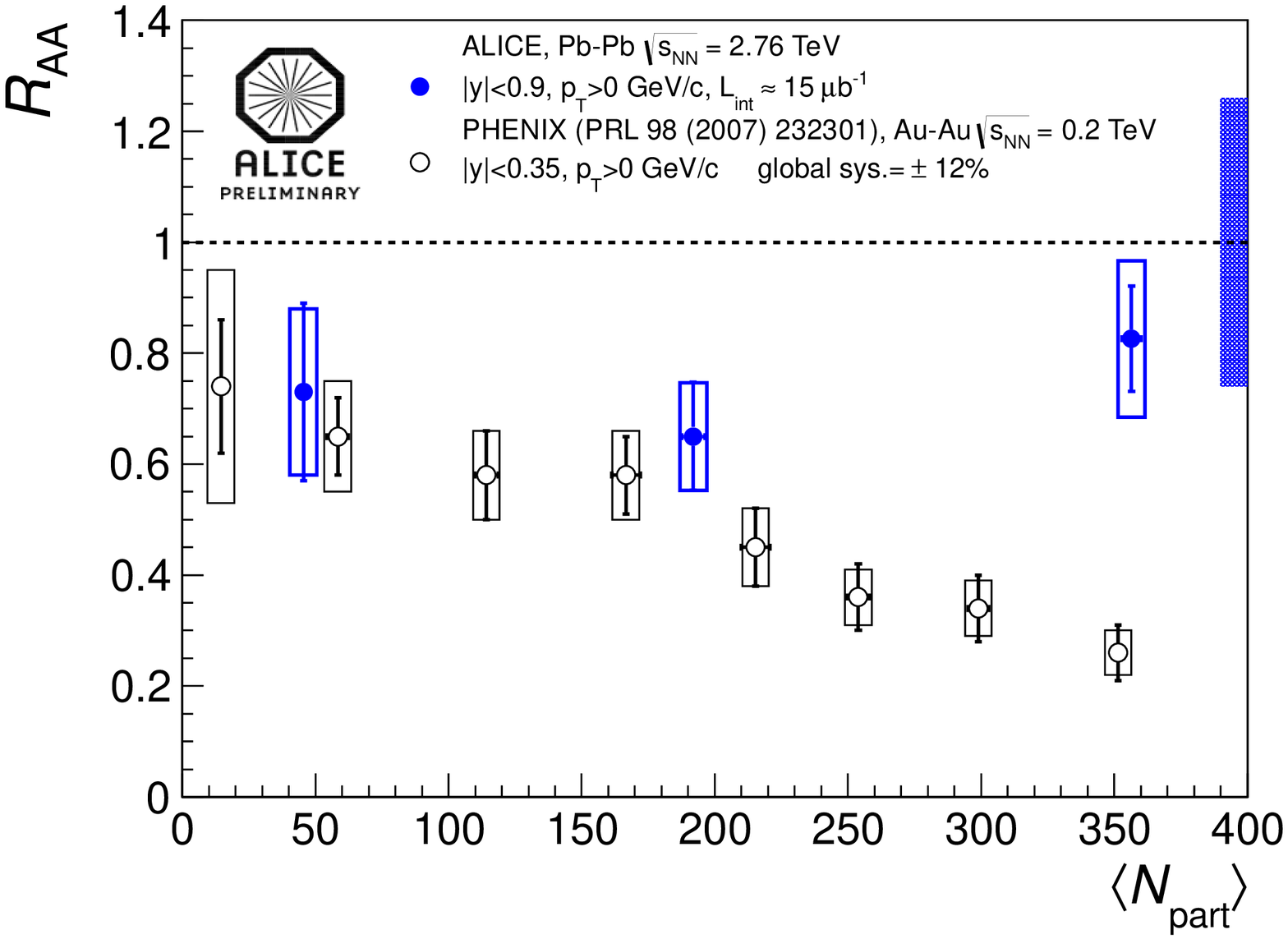}
&
\includegraphics[width=0.50\textwidth]{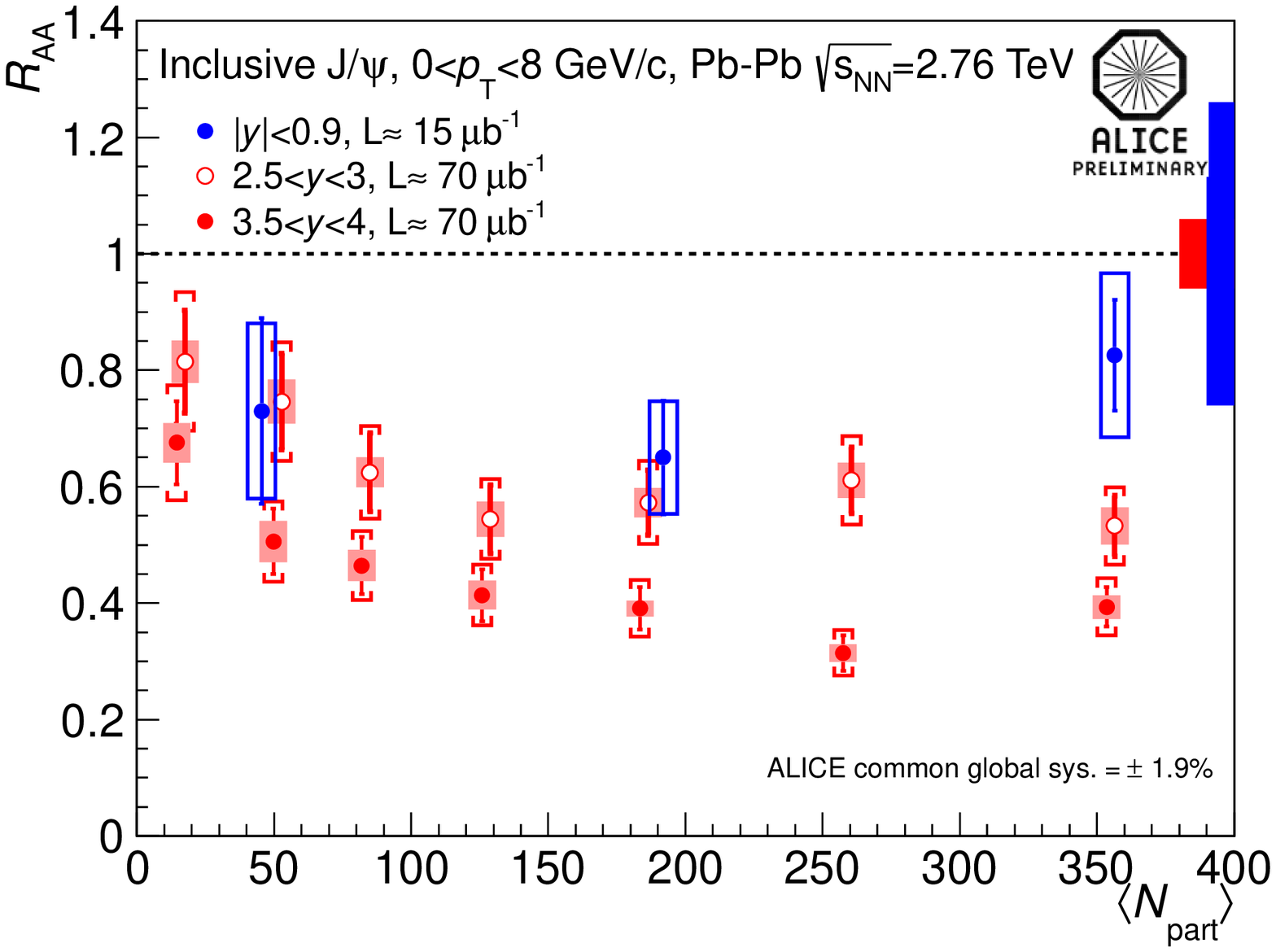} \\
\end{tabular}
\end{center}
\caption{Nuclear modification factor \raa as a function of the number of participants. 
         Comparison of the ALICE mid-rapidity \raa with the PHENIX mid-rapidity \raa (left) \cite{phenix2007}
         and the ALICE forward \raa (right) \cite{robertaQM2012}.}
\label{fig:results}
\end{figure}

The nuclear modification factor \raa for a given particle is defined as the ratio of the production rates
in nucleus-nucleus (AA) and nucleon-nucleon (NN) collisions normalized to the number of elementary collisions  (\ncoll)
in a nucleus-nucleus collision estimated from Glauber calculations. 
A trivial superposition of many elementary NN collisions would result in an \raa value equal to unity.
For the \jp we constructed the \raa using as reference the measured cross section
in pp collisions at $\sqrt{s}=$2.76~TeV \cite{jpsipp276}. The mid-rapidity inclusive \jp \raa as a function 
of the collision centrality is shown in Fig.\ref{fig:results}. The blue bar around 1 shown in both panels
indicate the total uncertainty from the pp reference. The error bars indicate the statistical uncertainty
while the boxes indicate the total systematic uncertainty of the Pb-Pb yields. The main contribution to the systematics
is coming from the signal extraction, other sources of uncertainty are the imperfect description
of the detector in the MC simulation, the kinematics of the input \jp spectrum used for corrections and the uncertainty on
\ncoll. 

\begin{figure}[!thbp]
\begin{center}
\centering
\includegraphics[width=0.60\textwidth]{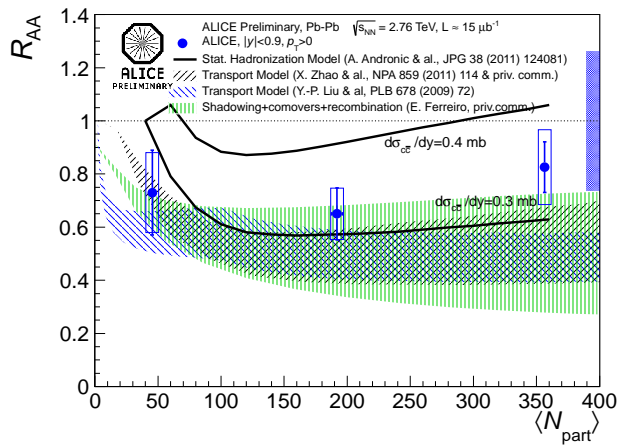}
\end{center}
\caption{\raa in Pb-Pb collisions at mid-rapidity as a function of centrality in comparison to several model predictions.}
\label{fig:models}
\end{figure}

For the most central 0-10\% collisions we obtained at mid-rapidity an \raa value of $0.83\pm0.09({\rm stat.})\pm0.26({\rm syst.})$ 
which is almost a factor 4 higher than the one measured by PHENIX in Au-Au collisions at \sq=200~GeV at mid-rapidity 
\cite{phenix2007}. Furthermore, our data indicate a small or no centrality dependence. The right-hand
side of Fig.\ref{fig:results} shows a comparison with the ALICE results at forward rapidity. The data indicate
that the \raa decreases with increasing rapidity but the large global systematic from the pp reference at mid-rapidity
prevents a very strong conclusion.

In Fig. \ref{fig:models} we show a comparison with theoretical models. All the models considered
here take into account the effect of (re)combination of charm quarks during the evolution of
the fireball or at freeze-out. Within the current experimental and theoretical uncertainties all the calculations 
describe the data.
The hashed bands show the results from two transport models \cite{zhao2011,liu2009} and from the comover
interaction model \cite{ferreiro}. In these models
the fraction of \jp's resulting from (re)combined \cc pairs is at most 50\%, the rest being produced
during the initial partonic stage of the collision. The solid lines show the prediction from the statistical hadronization 
model \cite{andronic2011}. This model assumes that 
no charmonium is formed in the QGP phase
and the charm quarks thermalize with the whole system. All the observed charmed hadrons 
are then formed at the freeze-out and their yields can be calculated based on the total charm cross section 
and the thermal model. All of the model calculations above have uncertainties due to the unknown 
$c\bar{c}$ production cross section. Nuclear shadowing also play an important role in the interpretation 
of these results. ALICE will measure this effect using the data in the LHC p-Pb run scheduled for beginning of 2013.

We reported on the latest ALICE measurements of the inclusive \jp nuclear modification factor as a function
of centrality in Pb-Pb collisions at \sq=2.76~TeV at mid-rapidity. The measured \raa indicates little or no
centrality dependence. In the most central collisions we observe
an \raa value which is almost a factor 4 higher than the results obtained in central Au-Au collisions by PHENIX.
Although the results have a large systematic uncertainty due to the pp reference at mid-rapidity, the data hints
that the \raa is decreasing with increasing rapidity. The comparison to the models considered here suggests 
that the predicted (re)combination effect plays an important role on the \jp production in Pb-Pb collisions
at LHC energies.

\end{document}